\documentclass[11pt,a4paper]{article}
\usepackage{amsmath,amsfonts,amssymb,amscd,epsfig,graphicx}
\catcode`\@=11
\@addtoreset{equation}{section}

\catcode`\@=12

\newfont{\gotico}{eufm10 scaled\magstephalf}
\newfont{\qvd}{msam10 scaled\magstephalf}

\def\de#1/de#2{\frac{\partial {#1}}{\partial {#2}}}
\def\De#1/de#2{\dfrac{\partial {#1}}{\partial {#2}}}

\begin{document}
\vskip-2cm
\title{Spin fluids in Bianchi-I $f(R)$-cosmology with torsion}
\author{Stefano Vignolo$^{1}$\footnote{E-mail: vignolo@diptem.unige.it} \ and
Luca Fabbri$^{1,2}$\footnote{E-mail: fabbri@diptem.unige.it}\\
\footnotesize{$^{1}$DIPTEM Sez. Metodi e Modelli Matematici, Universit\`{a} di Genova}\\
\footnotesize{Piazzale Kennedy, Pad. D - 16129 Genova (Italia)}\\
\footnotesize{$^{2}$INFN \& Dipartimento di Fisica, Universit\`{a} di Bologna}\\
\footnotesize{Via Irnerio 46, - 40126 Bologna (Italia)}}
\date{}
\maketitle
\begin{abstract}
\noindent We study Weyssenhoff spin fluids in Bianchi type-I cosmological models, within the framework of torsional $f(R)$-gravity; the resulting field equations are derived and discussed in both Jordan and Einstein frames, clarifying the role played by the spin and the non-linearity of the gravitational Lagrangian $f(R)$ in generating the torsional dynamical contributions. The general conservation laws holding for $f(R)$-gravity with torsion are employed to provide the conditions needed to ensure the preservation of the Hamiltonian constraint and the consequent correct formulation of the associated initial value problem. Examples are eventually given.
\end{abstract}
\section{Introduction}
Despite most of the success General Relativity (GR) enjoyed, there are some problems, at both ultraviolet and infrared scales, that GR is unlikely to solve, and therefore any of its enlargement is welcome: because the structure of the Einstein gravity is based on the assumption of least-order derivative field equations, by relaxing it means going to higher-order derivative models; these extensions however suffer from some arbitrariness, as nothing tells us what is the order derivative of the field equations.
\\
However, among the possible higher-order derivative field equations, or actions, there are two that are somehow special: one of them is based on the assumption of conformal invariance, and its beauty comes from the fact that this simple principle of symmetry is enough to severely restrict the possible actions to just a few; the other is based on the assumption of considering all gravitational information contained in the Ricci scalar curvature alone, then taking the most general form $f(R)$ for the gravitational action. Although the principle of conformal invariance is a beautiful and simple symmetry principle, nevertheless the universe does not display such a symmetry and a mechanism of spontaneous symmetry breaking must be invoked, and without it, it is preferable to investigate $f(R)$ gravity as a candidate to solve the problems we have mentioned above.
\\ 
In fact, $f(R)$ gravitation has been so far useful to address astrophysical and cosmological problems that we could otherwise faced only with the introduction of exotic form of matter and energy, the Dark Matter and Dark Energy; these forms of matter however are still a theoretical mystery. By converse, $f(R)$ gravitation has been able to account for both cosmic speed-up and missing matter without the introduction of new elements, as it is discussed in \cite{n-o,o,Capozziello:2011et}.
\\
On the other hand, the structure of the geometrical background must also take into account the matter fields we find in it: matter fields, according to the Wigner theoretical classification in terms of mass and spin, or the heuristic fact that they possess translational as well as rotational degrees of freedom, must have both energy and spin density; in a framework in which, as it appears to be reasonable, all physical quantities are coupled to geometrical ones, the energy is to be coupled to the curvature in the same spirit in which the spin is to be coupled to torsion. Thus we retain torsion not be neglected, likewise we would not neglect curvature, coupling the former to the spin, and the latter to the energy, and giving rise respectively to the Sciama-Kibble field equations, beside the Einstein field equations, as it has been discussed in \cite{hehl}.
\\
Now according to this imprinting, a theory of the $f(R)$ type in which the Ricci scalar curvature $R$ contains both metric and torsional degrees of freedom is the most straightforward extension we can take into account; an important consequence of the non-linearity of the function $f(R)$ is that we have non-vanishing torsion even without spin in general (that is if the energy-momentum trace is not constant \cite{CCSV1,CCSV2,CCSV3,CV4}). Then torsion may give rise to a repulsion ensuring singularity-free and accelerated cosmological models (\cite{SIV}).
\\
Besides, in a recent paper \cite{FV1} we have studied in detail this model in presence of Dirac fields, in order to better exploit the coupling between torsion and spin within the framework of $f(R)$-theories finding that, as it might have been somewhat expected, the angular degrees of freedom carried by the spin are amplified by its coupling to torsion in such a way that one is forced to deal at least with axial configurations in order to avoid consistency problems. Thus in another work \cite{VFC} we have focused on the Dirac field exploring cosmological applications for the simplest anisotropic universe given by the Bianchi type-I (BI).
\\
In the present work, we explore yet another situation, in which what is supposed to constitute the macroscopic approximation of the Dirac quantum field, the Weyssenhoff spin fluid, is studied in the context of BI universes.
\section{$f(R)$-gravity with torsion: generalities}
The dynamical gravitational variables of $f(R)$-gravity with torsion consist in pairs $(g,\Gamma)$, where $g$ is a metric tensor and $\Gamma$ is a metric-compatible linear connection. The field equations of the theory turn out to be \cite{CCSV1,CCSV2,CCSV3,CV4}
\begin{subequations}\label{2.1}
\begin{equation}
\label{2.1a}
f'\/(R)R_{ij} -\frac{1}{2}f\/(R)g_{ij}=\Sigma_{ij}
\end{equation}
\begin{equation}
\label{2.1b}
T_{ij}^{\;\;\;h}
=\frac{1}{f'(R)}
\left[\frac{1}{2}\left(\de{f'(R)}/de{x^{p}}+S_{pq}^{\;\;\;q}\right)
\left(\delta^{p}_{j}\delta^{h}_{i}-\delta^{p}_{i}\delta^{h}_{j}\right)
+S_{ij}^{\;\;\;h}\right]
\end{equation}
\end{subequations}
where $R_{ij}$ and $T_{ij}^{\;\;\;h}$ are the Ricci and Torsion tensors associated with $g$ and $\Gamma$, and $\Sigma_{ij}$ and $S_{ij}^{\;\;\;h}$ are the energy-momentum and spin density tensors of the matter fields. From \eqref{2.1b}, it is seen that we can have non-vanishing torsion even in absence of spin density. Following previous works \cite{CCSV1,CCSV2,CCSV3,CV4,FV1}, we suppose that the trace of equations \eqref{2.1a}
\begin{equation}\label{2.1bis}
f'(R)R -2f(R)=\Sigma
\end{equation}
gives rise to an invertible relation between the Ricci scalar curvature $R$ and the trace $\Sigma$ of the energy-momentum tensor. Also, we assume that $f(R)\not = kR^2$ (the case $f(R)=kR^2$ is only compatible with the condition $\Sigma=0$). Under the assumed conditions, from equation \eqref{2.1bis} it is possible to derive the expression of $R$ as a suitable function of $\Sigma$, namely
\begin{equation}\label{2.1tris}
R=F(\Sigma)
\end{equation}
Making use of the Bianchi identities, it is possible to derive the general conservation laws 
\begin{subequations}
\label{2.2}
\begin{equation}
\label{2.2a}
\nabla_{i}\Sigma^{ij}+T_{i}\Sigma^{ij}-\Sigma_{pi}T^{jpi}-\frac{1}{2}S_{sti}R^{stij}=0
\end{equation}
\begin{equation}
\label{2.2b}
\nabla_{h}S^{ijh}+T_{h}S^{ijh}+\Sigma^{ij}-\Sigma^{ji}=0
\end{equation}
\end{subequations}
under which the energy-momentum and spin density tensors of the matter fields must undergo once the matter field equations are assigned \cite{FV1}. In equations \eqref{2.2} the symbols $\nabla_i$ denote covariant derivative with respect to the dynamical connection $\Gamma$ while $R^{stij}$ is the curvature tensor of $\Gamma$. Indices are raised and lowered by the metric $g$.
\\
The dynamical connection can be expressed as
\begin{equation}\label{2.2bis}
\Gamma_{ij}^{\;\;\;h} = \tilde{\Gamma}_{ij}^{\;\;\;h} - K_{ij}^{\;\;\;h}
\end{equation}
where $\tilde{\Gamma}_{ij}^{\;\;\;h}$ denotes the Levi--Civita connection induced by the metric $g_{ij}$ and $K_{ij}^{\;\;\;h}$ indicates the contortion tensor \cite{Hehl}
\begin{equation}\label{2.3}
K_{ij}^{\;\;\;h} = \frac{1}{2}\/\left( - T_{ij}^{\;\;\;h} + T_{j\;\;\;i}^{\;\;h} - T^h_{\;\;ij}\right)
\end{equation}
Using the second set of field equations \eqref{2.2b}, we obtain then the following representations
\begin{subequations}\label{2.4}
\begin{equation}\label{2.4a}
K_{ij}^{\;\;\;h}= \hat{K}_{ij}^{\;\;\;h} + \hat{S}_{ij}^{\;\;\;h}
\end{equation}
\begin{equation}\label{2.4b}
\hat{S}_{ij}^{\;\;\;h}:=\frac{1}{2f'(R(\Sigma))}\/\left( - S_{ij}^{\;\;\;h} + S_{j\;\;\;i}^{\;\;h} - S^h_{\;\;ij}\right)
\end{equation}
\begin{equation}\label{2.4c}
\hat{K}_{ij}^{\;\;\;h} := -\hat{T}_j\delta^h_i + \hat{T}_pg^{ph}g_{ij}
\end{equation}
\begin{equation}\label{2.4d}
\hat{T}_j:=\frac{1}{2f'(R(\Sigma))}\/\left( \de{f'}/de{x^j} + S^\sigma_{j\sigma} \right)
\end{equation}
\end{subequations}
which, together with equation \eqref{2.2bis}, turn out to be very useful in order to separate the purely metric terms from the torsional ones in the field equations \eqref{2.1}.
\section{Coupling to spin fluids: field equations and conservation laws}
We consider a Weyssenhoff spin fluid, characterized by an energy-momentum tensor of the form
\begin{subequations}\label{2.5.1}
\begin{equation}\label{2.5.1a}
\Sigma^{ij}= U^iP^j + p\left( U^iU^j - g^{ij}\right)
\end{equation}
and a spin density tensor given by
\begin{equation}\label{2.5.1b}
S_{ij}^{\;\;\;h}=S_{ij}U^h
\end{equation}
\end{subequations}
where $U^i\/$ ($U^iU_i=1\/$) and $P^i$ denote respectively the $4$-velocity and the $4$-vector density of energy-momentum, while $S_{ij}\/$ the spin density of the fluid (see, for example, \cite{Obukhov,Hehl-Heyde-Kerlick} and references therein). The $4$-velocity and the spin density satisfy the convective condition
\begin{equation}\label{2.5.2}
S_{ij}U^j =0
\end{equation}
It is easily seen that the relations \eqref{2.5.2}, together with equation \eqref{2.4b}, imply the identities 
\begin{equation}\label{2.5.2bis}
\hat{S}_i^{\;\;ih}=-\hat{S}_i^{\;\;hi}=0
\end{equation}
Making use of \eqref{2.5.2bis} as well as of equations \eqref{2.2bis}, \eqref{2.3}, \eqref{2.4}, \eqref{2.5.1b} and \eqref{2.5.2}, we can express the contracted curvature and the scalar curvature respectively as
\begin{subequations}\label{2.5.3}
\begin{equation}\label{2.5.3a}
\begin{split}
R_{ij} = \tilde{R}_{ij} - 2\tilde{\nabla}_{j}\hat{T}_i - \tilde{\nabla}_h\hat{T}^hg_{ij} + 2\hat{T}_i\hat{T}_j - 2\hat{T}_h\hat{T}^hg_{ij} - \frac{1}{f'}\hat{T}_hS^h_{\;\;j}U_i \\
- \frac{1}{2f'}\tilde{\nabla}_h\/\left(- S_{ji}U^h + S_i^{\;\;h}U_j - S^h_{\;\;j}U_i\right) + \frac{1}{4(f')^2}S^{pq}S_{pq}U_iU_j
\end{split}
\end{equation}
and
\begin{equation}\label{2.5.3b}
R = \tilde{R} - 6\tilde{\nabla}_{i}\hat{T}^i - 6\hat{T}_i\hat{T}^i + \frac{1}{4(f')^2}S^{pq}S_{pq}
\end{equation}
\end{subequations}
where now $\hat{T}_i =\frac{1}{2f'}\de{f'}/de{x^i}\/$. In view of this, substituting equations \eqref{2.5.3} in equations \eqref{2.1}, we obtain Einstein-like equations of the form
\begin{equation}\label{2.5.4}
\begin{split}
\tilde{R}_{ij} -\frac{1}{2}\tilde{R}g_{ij}= \frac{1}{\varphi}\Sigma_{ij} + \frac{1}{\varphi^2}\left( - \frac{3}{2}\de\varphi/de{x^i}\de\varphi/de{x^j} + \varphi\tilde{\nabla}_{j}\de\varphi/de{x^i} + \frac{3}{4}\de\varphi/de{x^h}\de\varphi/de{x^k}g^{hk}g_{ij} \right. \\
\left. - \varphi\tilde{\nabla}^h\de\varphi/de{x^h}g_{ij} - V\/(\varphi)g_{ij} \right) + \frac{1}{\varphi}\hat{T}_hS^h_{\;\;j}U_i + \frac{1}{2\varphi}\tilde{\nabla}_h\/\left(- S_{ji}U^h + S_i^{\;\;h}U_j - S^h_{\;\;j}U_i\right) \\
- \frac{1}{4\varphi^2}S^{pq}S_{pq}U_iU_j + \frac{1}{8\varphi^2}S^{pq}S_{pq}g_{ij}
\end{split}
\end{equation}
where, using equation \eqref{2.1tris}, the scalar function
\begin{equation}\label{2.5.7}
\varphi := f'\/(F\/(\Sigma))
\end{equation}
and the effective potential \cite{CCSV1}
\begin{equation}\label{2.5.8}
V\/(\varphi):= \frac{1}{4}\left[ \varphi F^{-1}\/((f')^{-1}\/(\varphi)) + \varphi^2\/(f')^{-1}\/(\varphi)\right]
\end{equation}
have been introduced. It is a straightforward matter to verify that the conservation laws for the spin \eqref{2.2b} amount to the validity of the anti-symmetrized part of the Einstein-like equations \eqref{2.5.4}
\begin{equation}\label{2.5.9}
\tilde{\nabla}_h\/\left(S_{ij}U^h \right) + \hat{T}_hS^h_{\;\;j}U_i - \hat{T}_hS^h_{\;\;i}U_j + \Sigma_{ij} - \Sigma_{ji} =0
\end{equation}
Saturating equations \eqref{2.5.9} with $U^i$ we obtain the explicit expression for the $4$-vector density of energy-momentum
\begin{equation}\label{2.5.10}
P_j = \rho\,U_j - \hat{T}_h\/S^h_{\;\;j} - \tilde{\nabla}_h\/\left(S_{ij}U^h \right)U^i
\end{equation}
where $\rho := U^iP_i$. inserting equations \eqref{2.5.10} into \eqref{2.5.1a} we get the explicit form of the energy-momentum tensor
\begin{equation}\label{2.5.11}
\Sigma_{ij} = (\rho + p)\,U_iU_j - p\,g_{ij} - U_i\hat{T}_h\/S^h_{\;\;j} - U_i\tilde{\nabla}_h\/\left(S_{kj}U^h \right)U^k
\end{equation}
The significant part of the Einstein-like equations is then the symmetric one
\begin{equation}\label{2.5.12}
\begin{split}
\tilde{R}_{ij} -\frac{1}{2}\tilde{R}g_{ij}= \frac{1}{\varphi}\Sigma_{(ij)} + \frac{1}{\varphi^2}\left( - \frac{3}{2}\de\varphi/de{x^i}\de\varphi/de{x^j} + \varphi\tilde{\nabla}_{j}\de\varphi/de{x^i} + \frac{3}{4}\de\varphi/de{x^h}\de\varphi/de{x^k}g^{hk}g_{ij} \right. \\
\left. - \varphi\tilde{\nabla}^h\de\varphi/de{x^h}g_{ij} - V\/(\varphi)g_{ij} \right) + \frac{1}{2\varphi}\hat{T}_hS^h_{\;\;j}U_i + \frac{1}{2\varphi}\hat{T}_hS^h_{\;\;i}U_j+ \frac{1}{2\varphi}\tilde{\nabla}_h\/\left( - S^h_{\;\;i}U_j - S^h_{\;\;j}U_i\right) \\
- \frac{1}{4\varphi^2}S^{pq}S_{pq}U_iU_j + \frac{1}{8\varphi^2}S^{pq}S_{pq}g_{ij}
\end{split}
\end{equation}
Moreover, working out the conservation laws \eqref{2.2a}, the latter are seen to assume the form
\begin{equation}\label{2.5.13}
\begin{split}
\tilde{\nabla}_i\Sigma^{(ij)} + \tilde{\nabla}_i\left(-\frac{1}{4\varphi}S_{hk}S^{hk}U^iU^j\right) - \frac{1}{2\varphi}\hat{T}^jS_{hi}S^{hi} + \frac{1}{8\varphi}\tilde{\nabla}^j\left(S_{hi}S^{hi}\right) +\\
+ S^{hi}\tilde{\nabla}_h\tilde{\nabla}_iU^j - \frac{1}{2}\tilde{\nabla}_i\tilde{\nabla}_h\left(S^{ij}U^h\right) + \tilde{\nabla}_i\left(\frac{1}{2}\hat{T}_hS^{hj}U^i + \frac{1}{2}\hat{T}_hS^{hi}U^j\right) =0
\end{split}
\end{equation}
Equations \eqref{2.5.13} allow to clarify the relationship between the conservation laws \eqref{2.2} and the symmetrized Einstein-like equations \eqref{2.5.12}. To start with, we recall that equations \eqref{2.5.7} and \eqref{2.5.8} are equivalent to the relation 
\begin{equation}\label{A.9}
\Sigma -\frac{6}{\varphi}V(\varphi) + 2V'(\varphi)=0
\end{equation}
(see \cite{CCSV1} for the proof). After that, taking the trace of equations \eqref{2.5.12}, we get 
\begin{equation}\label{A.10}
\Sigma= -\varphi\tilde{R} - \frac{3}{2\varphi}\varphi_i\varphi^i + 3\tilde{\nabla}_i\varphi^i + \frac{4}{\varphi}V(\varphi) - \frac{1}{4\varphi}S_{pq}S^{pq}
\end{equation}
Substituting equation \eqref{A.10} in equation \eqref{A.9}, we obtain
\begin{equation}\label{A.11}
\tilde{R} + \frac{3}{2\varphi^2}\varphi_i\varphi^i - \frac{3}{\varphi}\tilde{\nabla}_i\varphi^i + 
\frac{2}{\varphi^2}V(\varphi) - \frac{2}{\varphi}V'(\varphi) + \frac{1}{4\varphi^2}S_{pq}S^{pq}=0
\end{equation}
We rewrite equation \eqref{2.5.12} in the form
\begin{equation}\label{A.12}
\begin{split}
\varphi\tilde{R}_{ij} -\frac{\varphi}{2}\tilde{R}g_{ij}= \Sigma_{(ij)} + \frac{1}{\varphi}\left( - \frac{3}{2}\de\varphi/de{x^i}\de\varphi/de{x^j} + \varphi\tilde{\nabla}_{j}\de\varphi/de{x^i} + \frac{3}{4}\de\varphi/de{x^h}\de\varphi/de{x^k}g^{hk}g_{ij} \right. \\
\left. - \varphi\tilde{\nabla}^h\de\varphi/de{x^h}g_{ij} - V\/(\varphi)g_{ij} \right) + \frac{1}{2}\hat{T}_hS^h_{\;\;j}U_i + \frac{1}{2}\hat{T}_hS^h_{\;\;i}U_j+ \frac{1}{2}\tilde{\nabla}_h\/\left( - S^h_{\;\;i}U_j - S^h_{\;\;j}U_i\right) \\
- \frac{1}{4\varphi}S^{pq}S_{pq}U_iU_j + \frac{1}{8\varphi}S^{pq}S_{pq}g_{ij}
\end{split}
\end{equation}
The covariant divergence of \eqref{A.12} yields
\begin{equation}\label{A.13}
\begin{split}
(\tilde\nabla^i\varphi)\tilde{R}_{ij} + \varphi\tilde\nabla^i\tilde{G}_{ij} -\frac{1}{2}\tilde{R}\tilde\nabla_j\varphi = \tilde\nabla^i\Sigma_{(ij)} + \left(\tilde\nabla^i\tilde{\nabla}_{i}\tilde\nabla_j - \tilde\nabla_j\tilde{\nabla}^i\tilde\nabla_i\right)\varphi +\\
+ \tilde\nabla^i\left[\frac{1}{\varphi}\left(-\frac{3}{2}\varphi_i\varphi_j + \frac{3}{4}\varphi_h\varphi^h\/g_{ij} - V\/(\varphi)g_{ij}\right)\right] + \tilde{\nabla}^i\left(\frac{1}{2}\hat{T}_hS^h_{\;\;j}U_i + \frac{1}{2}\hat{T}_hS^h_{\;\;i}U_j\right) +\\
-\frac{1}{2}\tilde{\nabla}^i\tilde{\nabla}_h\left(S^h_{\;\;i}U_j + S^h_{\;\;j}U_i\right) - \frac{1}{4}\tilde{\nabla}^i\left(\frac{1}{\varphi}S^{pq}S_{pq}U_iU_j\right) + \frac{1}{8}\tilde{\nabla}^i\left(\frac{1}{\varphi}S^{pq}S_{pq}g_{ij}\right)
\end{split}
\end{equation}
By definition, the Einstein and the Ricci tensors satisfy $\tilde\nabla^j\tilde{G}_{ij}=0$ and $(\tilde\nabla^j\varphi)\tilde{R}_{ij} = \left(\tilde\nabla^j\tilde{\nabla}_{j}\tilde\nabla_i - \tilde\nabla_i\tilde{\nabla}^j\tilde\nabla_j\right)\varphi$. Then equation \eqref{A.13} reduces to
\begin{equation}\label{A.14}
\begin{split}
-\frac{1}{2}\tilde{R}\tilde\nabla_j\varphi = \tilde\nabla^i\tilde\Sigma_{(ij)}+ \tilde\nabla^i\left[\frac{1}{\varphi}\left(-\frac{3}{2}\varphi_i\varphi_j + \frac{3}{4}\varphi_h\varphi^h\/g_{ij} - V\/(\varphi)g_{ij}\right)\right]+\\
+ \tilde{\nabla}^i\left(\frac{1}{2}\hat{T}_hS^h_{\;\;j}U_i + \frac{1}{2}\hat{T}_hS^h_{\;\;i}U_j\right) - \frac{1}{2}\tilde{\nabla}^i\tilde{\nabla}_h\left(S^h_{\;\;i}U_j + S^h_{\;\;j}U_i\right) +\\
- \frac{1}{4}\tilde{\nabla}^i\left(\frac{1}{\varphi}S^{pq}S_{pq}U_iU_j\right) + \frac{1}{8}\tilde{\nabla}^i\left(\frac{1}{\varphi}S^{pq}S_{pq}g_{ij}\right)
\end{split} 
\end{equation}
On the other hand, making use of equation \eqref{A.11} it is easily seen that
\begin{equation}\label{A.15}
-\frac{1}{2}\tilde{R}\tilde\nabla_j\varphi = \tilde\nabla^i\left[\frac{1}{\varphi}\left(-\frac{3}{2}\varphi_i\varphi_j + \frac{3}{4}\varphi_h\varphi^h\/g_{ij} - V\/(\varphi)g_{ij}\right)\right] - \tilde{\nabla}_j\left(\frac{1}{8\varphi}\right)S_{hi}S^{hi}
\end{equation}
Inserting equations \eqref{2.5.13} and \eqref{A.15} into \eqref{A.14}, we end up with final equations
\begin{equation}\label{A.16}
- S^{hi}\tilde{\nabla}_h\tilde{\nabla}_iU_j + \frac{1}{2}\tilde{\nabla}_i\tilde{\nabla}_h\left(S^{i}_{\;\;j}U^h\right) - \frac{1}{2}\tilde{\nabla}^i\tilde{\nabla}_h\left(S^h_{\;\;i}U_j + S^h_{\;\;j}U_i\right) =0
\end{equation}
representing the requirement of vanishing for the covariant divergence of the symmetrized Einstein-like equations, written in the form \eqref{A.12}. As we shall see in the next section, equations \eqref{A.16} have to be imposed in order to achieve consistency between the assignment of initial data, their preservation in time and the field equations.
\section{Bianchi-I cosmological models}
In this section we shall study Bianchi type-I models arising from the above described theory. To this end, we consider a Bianchi type-I metric 
\begin{equation}\label{4.1}
ds^2 = dt^2 - a^2(t)\,dx^2 - b^2(t)\,dy^2 - c^2(t)\,dz^2
\end{equation}
The non-trivial Christoffel symbols associated with the metric \eqref{4.1} are
\begin{equation}\label{4.2}
\begin{split}
\tilde{\Gamma}_{10}^{\;\;\;1}= \frac{\dot a}{a}, \quad \tilde{\Gamma}_{20}^{\;\;\;2}= \frac{\dot b}{b}, \quad \tilde{\Gamma}_{30}^{\;\;\;3}= \frac{\dot c}{c}\\
\tilde{\Gamma}_{11}^{\;\;\;0}= a{\dot a}, \quad \tilde{\Gamma}_{22}^{\;\;\;0}= b{\dot b}, \quad \tilde{\Gamma}_{33}^{\;\;\;0}= c{\dot c}
\end{split}
\end{equation}
Setting $U^i=(1,0,0,0)$, the convective condition \eqref{2.5.2} implies $S_{0i}=0$; from this, making use of equations \eqref{4.2} and taking the hypothesis of homogeneity into account, it is easily seen that the energy-momentum tensor reduces to
\begin{equation}\label{4.2bisbis}
\Sigma_{ij} = (\rho + p)\,U_iU_j - p\,g_{ij}
\end{equation}
as well as that the non-diagonal part of the Einstein-like equations \eqref{2.5.12}, evaluated for the metric \eqref{4.1}, yields the equations
\begin{subequations}\label{4.2bis}
\begin{equation}\label{4.2bisa}
S_{12}\left(\frac{\dot a}{a} - \frac{\dot b}{b}\right)=0
\end{equation}
\begin{equation}\label{4.2bisb}
S_{13}\left(\frac{\dot a}{a} - \frac{\dot c}{c}\right)=0
\end{equation}
\begin{equation}\label{4.2bisc}
S_{23}\left(\frac{\dot b}{b} - \frac{\dot c}{c}\right)=0
\end{equation}
\end{subequations}
Neglecting the trivial case (zero spin and total isotropy), equations \eqref{4.2bis} admit three kinds of possible solutions: the first one is given by $S_{ij}=0$ (zero spin); the second one is given by $\frac{\dot a}{a}=\frac{\dot b}{b}=\frac{\dot c}{c}$ (total isotropy); the third one is characterized by a partial isotropy and by a spin aligned along a spatial axis. It is clear that the only significant solutions are represented by the third case. Indeed, on the one hand in the first case we would have an anisotropic universe without spin density, thus the anisotropy would be unjustified; on the other hand, in the second and opposite case, we would have discrepancy between the geometrical isotropy of the universe and the non-vanishing spin density generating a spin vector with non-zero spatial components. Therefore, in the subsequent discussion, we shall focus on solutions of the third kind only. For instance, supposing the spin aligned along the $z$-axis, a Bianchi type-I metric consistent with the stated assumption is then of the form \eqref{4.1} with $a(t)=b(t)$, while $S_{12}=-S_{21}$ are the only non-vanishing components of the spin tensor. In such a circumstance, the diagonal part of the Einstein-like equations gives rise to the equations
\begin{subequations}\label{4.3}
\begin{equation}\label{4.3a}
\left(\frac{\dot a}{a}\right)^2 + 2\frac{\dot a}{a}\frac{\dot c}{c} = \frac{\rho}{\varphi} + 
\frac{1}{\varphi^2}\left[- \frac{3}{4}{\dot\varphi}^2 - \varphi\dot\varphi\frac{\dot\tau}{\tau} - V(\varphi)\right] - \frac{1}{4\varphi^2}S^2
\end{equation}
\begin{equation}\label{4.3b}
\frac{\ddot a}{a} + \frac{\ddot c}{c} + \frac{\dot a}{a}\frac{\dot c}{c} = - \frac{p}{\varphi} + 
\frac{1}{\varphi^2}\left[\varphi\dot\varphi\frac{\dot a}{a} + \frac{3}{4}{\dot\varphi}^2 -\varphi\left( \ddot\varphi + \frac{\dot\tau}{\tau}\dot\varphi \right) - V(\varphi)\right] + \frac{1}{4\varphi^2}S^2
\end{equation}
\begin{equation}\label{4.3c}
2\frac{\ddot a}{a} + \left(\frac{\dot a}{a}\right)^2 = - \frac{p}{\varphi} + 
\frac{1}{\varphi^2}\left[\varphi\dot\varphi\frac{\dot c}{c} + \frac{3}{4}{\dot\varphi}^2 -\varphi\left( \ddot\varphi + \frac{\dot\tau}{\tau}\dot\varphi \right) - V(\varphi)\right] + \frac{1}{4\varphi^2}S^2
\end{equation}
\end{subequations}
where we have put $\tau:= a^2c$ and $S^2 := \frac{1}{2}S^{pq}S_{pq}$ \cite{s}. In this case, the conservation laws \eqref{2.5.9} and \eqref{2.5.13} give rise to the equations
\begin{subequations}\label{4.4}
\begin{equation}\label{4.4a}
\dot\rho + \left(\rho + p\right)\frac{\dot\tau}{\tau}=0
\end{equation}
\begin{equation}\label{4.4b}
\dot{S} + S\frac{\dot\tau}{\tau}=0
\end{equation}
\end{subequations}
Supposing an equation of state of the kind $p=\lambda\rho$ ($0\leq\lambda\leq 1$), the latter admit general solutions
\begin{equation}\label{4.5}
\rho = \rho_0\tau^{-(1+\lambda)}, \qquad S=\frac{S_0}{\tau}
\end{equation}
$\rho_0$ and $S_0$ being suitable integration constants. Moreover, under the stated hypotheses on the metric tensor and the spin tensor, it is an easy matter to verify that equations \eqref{A.16} are trivial identities automatically satisfied.
\\
Concerning the Einstein-like equations \eqref{4.3}, we can subtract equation \eqref{4.3b} from equation \eqref{4.3c} obtaining
\begin{equation}\label{4.6}
\varphi\tau\frac{d}{dt}\left(\frac{\dot a}{a} - \frac{\dot c}{c}\right) + \varphi\dot\tau\left(\frac{\dot a}{a} - \frac{\dot c}{c}\right) + \dot{\varphi}\tau\left(\frac{\dot a}{a} - \frac{\dot c}{c}\right)=0
\end{equation}
from which we derive
\begin{equation}\label{4.7}
\frac{a}{c}=De^{\left(X\int{\frac{dt}{\varphi\tau}}\right)}
\end{equation}
$D$ and $X$ being suitable constants, which gives the evolution equation for the ratio of the two axes that are not constantly proportional, and therefore it can be considered as the evolution equation for the shape of the universe; from equation \eqref{4.7} the two explicit expressions follow
\begin{subequations}\label{4.7bis}
\begin{equation}\label{4.7bisa}
a=D^{\frac{1}{3}}\tau^{\frac{1}{3}}e^{\left(\frac{X}{3}\int{\frac{dt}{\varphi\tau}}\right)}
\end{equation}
\begin{equation}\label{4.7bisb}
c=D^{-\frac{2}{3}}\tau^{\frac{1}{3}}e^{\left(\frac{-2X}{3}\int{\frac{dt}{\varphi\tau}}\right)}
\end{equation}
\end{subequations}
Also, multiplying \eqref{4.3a} by $3$ and adding the result to the summation of two times equation \eqref{4.3b} with \eqref{4.3c}, we get the final equation for $\tau$
\begin{equation}\label{4.8}
2\frac{\ddot\tau}{\tau} + 3\frac{\ddot\varphi}{\varphi} = 3\frac{\rho}{\varphi} - 3\frac{p}{\varphi} - 5\frac{\dot\tau}{\tau}\frac{\dot\varphi}{\varphi} - \frac{6}{\varphi^2}V\/(\varphi)
\end{equation}
which can be considered as the evolution equation of the volume of the universe. Here, it is worth noticing that in equation \eqref{4.8} (as well as in equation \eqref{4.6}) there is no longer the presence of spin contributions. Another relevant remark is that equation \eqref{4.3a} plays the role of a constraint on the initial data: thus for consistency we have to check that, if satisfied initially, this constraint is preserved in time. To see this point, we first observe that the Einstein-like equations \eqref{2.5.12} (or \eqref{A.12}), and thus also \eqref{4.3}, can be written in the equivalent form
\begin{equation}\label{3.14.1}
\tilde{R}_{ij}= \tilde{T}_{ij} -\frac{1}{2}\tilde{T}g_{ij}
\end{equation}
where
\begin{equation}\label{3.14.2}
\begin{split}
\tilde{T}_{ij}:= \frac{1}{\varphi}\Sigma_{(ij)} + \frac{1}{\varphi^2}\left( - \frac{3}{2}\de\varphi/de{x^i}\de\varphi/de{x^j} + \varphi\tilde{\nabla}_{j}\de\varphi/de{x^i} + \frac{3}{4}\de\varphi/de{x^h}\de\varphi/de{x^k}g^{hk}g_{ij} \right. \\
\left. - \varphi\tilde{\nabla}^h\de\varphi/de{x^h}g_{ij} - V\/(\varphi)g_{ij} \right) + \frac{1}{2\varphi}\hat{T}_hS^h_{\;\;j}U_i + \frac{1}{2\varphi}\hat{T}_hS^h_{\;\;i}U_j+ \frac{1}{2\varphi}\tilde{\nabla}_h\/\left( - S^h_{\;\;i}U_j - S^h_{\;\;j}U_i\right) \\
- \frac{1}{4\varphi^2}S^{pq}S_{pq}U_iU_j + \frac{1}{8\varphi^2}S^{pq}S_{pq}g_{ij}
\end{split}
\end{equation}
denotes the effective energy-momentum tensor appearing on the right hand side of equations \eqref{2.5.12}, while $\tilde T$ is its trace. It is then a straightforward matter to verify that equations \eqref{4.6} and \eqref{4.8} can be equivalently obtained by suitably combining the space-space equations of the set \eqref{3.14.1}; more in detail, equation \eqref{4.6} is obtained by subtracting form one another the two distinct space-space equations of \eqref{3.14.1}, while equation \eqref{4.8} is obtained adding together all the space-space equations of \eqref{3.14.1}. As a consequence, we have that solving equations \eqref{4.6} and \eqref{4.8} amounts to solve all the space-space equations of the set \eqref{3.14.1}. In addition to this, as it is proved in Section 3, the conservation laws \eqref{2.2} together with equations \eqref{A.16} automatically imply the vanishing of the four-divergence of the Einstein-like equations in the form \eqref{A.12}. The two just mentioned facts allow to apply to the present case a result by Bruhat (see \cite{yvonne4}, Theorem 4.1, pag. 150) which ensures that the constraint \eqref{4.3a} is actually satisfied for all time.
\section{From the Jordan to the Einstein frame}
In this section we shall show that, by passing from the Jordan frame ($g_{ij}$) to the Einstein frame ($\bar{g}_{ij}$) through the conformal transformation $\bar{g}_{ij}:=\varphi\/g_{ij}$ (if $\varphi>0$) or $\bar{g}_{ij}:=-\varphi\/g_{ij}$ (if $\varphi<0$), we can obtain a final differential equation for the volume-scale of the universe which is simpler than equation \eqref{4.8} and, in general, can be quadratically integrable. The new equation arising in the Einstein frame allows a qualitative analysis of the corresponding solutions even when it is difficult to solve analytically. 
\\
To achieve our aim, we have to perform two distinct steps: the first one consists in expressing all field equations \eqref{2.5.9}, \eqref{2.5.12}, \eqref{2.5.13} and \eqref{A.16} in terms of the conformal metric $\bar{g}_{ij}$, its Levi--Civita connection $\bar{\Gamma}_{ij}^{\;\;\;h}$ and the four velocity $\bar{U}^i$ ($\bar{g}_{ij}\bar{U}^i\bar{U}^j=1$); the second one is evaluating all the so obtained field equations for a Bianchi type-I solution of the kind
\begin{equation}\label{C.1}
d\bar{s}^2 = dt^2 - \bar{a}^2(t)\,\left(dx^2 + dy^2\right) - \bar{c}^2(t)\,dz^2
\end{equation}  
and for a spin tensor whose unique non-vanishing components are again $S_{12}=-S_{21}$ (spin aligned along the $z$-axis).
From now on, for simplicity, we suppose that $\varphi>0$. Introducing the conformal metric $\bar{g}_{ij}=\varphi\/g_{ij}$, it is an easy matter to verify that
\begin{equation}\label{5.1}
\bar{G}_{ij} = \tilde{G}_{ij} - \frac{1}{\varphi^2}\left[ - \frac{3}{2}\varphi_i\varphi_j + \varphi\tilde{\nabla}_{j}\varphi_i + \frac{3}{4}\varphi_h\varphi^h\/g_{ij} - \varphi(\tilde{\nabla}^h\varphi_h)g_{ij} \right]
\end{equation}
where $\bar{G}_{ij}$ and $\tilde{G}_{ij}$ are the Einstein tensors associated with the metrics $\bar{g}_{ij}$ and $g_{ij}$ respectively. We can then write the symmetrized Einstein-like equations \eqref{2.5.12} in the simpler form
\begin{equation}\label{5.2}
\begin{split}
\bar{G}_{ij} = \frac{1}{\varphi}\Sigma_{(ij)} - \frac{1}{\varphi^2}V\/(\varphi)g_{ij} + \frac{1}{2\varphi}\left(\hat{T}_hS^h_{\;\;j}U_i + \hat{T}_hS^h_{\;\;i}U_j\right) + \\
+ \frac{1}{2\varphi}\tilde{\nabla}_h\/\left( - S^h_{\;\;i}U_j - S^h_{\;\;j}U_i\right) - \frac{1}{4\varphi^2}S^{pq}S_{pq}U_iU_j + \frac{1}{8\varphi^2}S^{pq}S_{pq}g_{ij}
\end{split}
\end{equation}
Equations \eqref{5.2} are the key point of our discussion. To start with, making use of the relation 
\begin{equation}\label{B.1}
\bar{\Gamma}_{ij}^{\;\;\;h}= \tilde{\Gamma}_{ij}^{\;\;\;h} + \frac{1}{2\varphi}\varphi_j\delta^h_i - \frac{1}{2\varphi}\varphi_p\/g^{ph}g_{ij} + \frac{1}{2\varphi}\varphi_i\delta^h_j
\end{equation}
linking the Levi--Civita connections associated with the metric tensors $\bar{g}_{ij}$ and $g_{ij}$ respectively, and taking the identities $U^i = \sqrt{\varphi}\bar{U}^i$ ($U_i = \frac{1}{\sqrt{\varphi}}\bar{U}_i$) into account, it is easily seen that under the stated hypotheses the energy-momentum tensor \eqref{2.5.11} assumes the form
\begin{equation}\label{C.2}
\Sigma_{ij}:= \frac{1}{\varphi}(\rho +p)\/\bar{U}_i\bar{U}_j -\frac{p}{\varphi}\bar{g}_{ij}
\end{equation}
Inserting the content of equations \eqref{C.1}, \eqref{B.1} and \eqref{C.2} into equations \eqref{5.2}, we get 
\begin{subequations}\label{C.3}
\begin{equation}\label{C.3a}
\left(\frac{\dot{\bar a}}{\bar a}\right)^2 + 2\frac{\dot{\bar a}}{\bar a}\frac{\dot{\bar c}}{\bar c} = \frac{\rho}{\varphi^2} - \frac{1}{\varphi^3}V(\varphi) - \frac{1}{4\varphi^3}S^2 
\end{equation}
and
\begin{equation}\label{C.3b}
\frac{\ddot{\bar a}}{\bar a} + \frac{\ddot{\bar c}}{\bar c} + \frac{\dot{\bar a}}{\bar a}\frac{\dot{\bar c}}{\bar c} = - \frac{p}{\varphi^2} - 
\frac{1}{\varphi^3}V(\varphi) + \frac{1}{4\varphi^3}S^2
\end{equation}
\begin{equation}\label{C.3c}
2\frac{\ddot{\bar a}}{\bar a}  + \left(\frac{\dot{\bar a}}{\bar a}\right)^2= - \frac{p}{\varphi^2} - 
\frac{1}{\varphi^3}V(\varphi) + \frac{1}{4\varphi^3}S^2
\end{equation}
\end{subequations}
In a similar way, from the conservation laws \eqref{2.5.9} and \eqref{2.5.13} we obtain the equations
\begin{equation}\label{C.4}
\frac{\dot S}{S} + \frac{\dot{\bar\tau}}{\bar\tau} - \frac{3}{2}\frac{\dot\varphi}{\varphi}
\end{equation} 
and
\begin{equation}\label{C.5}
\frac{\dot\rho}{\rho} + (1+\lambda)\frac{\dot{\bar\tau}}{\bar\tau} - \frac{3}{2}\frac{\dot\varphi}{\varphi}(1+\lambda)=0
\end{equation}
Setting $\Theta:= \pm\varphi^{-\frac{3}{2}}$, one has $- \frac{3}{2}\frac{\dot\varphi}{\varphi}=\frac{\dot\Theta}{\Theta}$ 
and then we can integrate equation \eqref{C.4} and \eqref{C.5} as
\begin{equation}\label{C.6}
S=\frac{S_0}{\bar\tau\Theta}, \qquad \rho =\rho_0\/(\bar{\tau}\Theta)^{-(1+\lambda)}
\end{equation}
where $S_0$ and $\rho_0$ are integration constants. Concerning equations \eqref{C.3}, we can proceed as in the Jordan frame (see section 4), so obtaining the relations 
\begin{equation}\label{C.7}
\frac{\bar a}{\bar c} = D\exp\left(X\int{\frac{1}{\bar\tau}dt}\right)
\end{equation}
and 
\begin{equation}\label{C.8}
2\frac{\ddot{\bar\tau}}{\bar\tau} = 3\frac{\rho}{\varphi^2} - 3\frac{p}{\varphi^2} - \frac{6}{\varphi^3}V\/(\varphi)
\end{equation}
where $D$ and $X$ denote suitable integration constants. We stress that, as in equation \eqref{4.8}, also in equation \eqref{C.8} there are no spin contributions. At this point, we notice that if we could express the quantities $\rho$, and $\varphi$ (or, equivalently, $\Theta=\varphi^{-\frac{3}{2}}$) as functions of $\bar\tau$ and insert the so obtained results into equation \eqref{C.8}, we would get a final equation for $\bar\tau$ of the kind 
\begin{equation}\label{C.9}
\ddot{\bar\tau}= f(\bar\tau)
\end{equation}
which is quadratically integrable. The second equation of \eqref{C.6} together with $\Theta^{-\frac{2}{3}}=\varphi=f'(R(\rho))$
constitute a set of two relations involving the three variables $\bar\tau$, $\rho$ e $\Theta$. In principle, it is then possible to express the last two variables as suitable functions of $\bar\tau$ alone, thus obtaining from \eqref{C.8} a final quadratically integrable equation of the kind \eqref{C.9}.
\\
As a last remark, we note that equation \eqref{C.3a} has the nature of a constraint on the initial data. As made in the Jordan frame, we have again to check that such a constraint is satisfied for all time after the initial instant. In connection with this, we now prove that the conservation laws \eqref{2.2} together with equations \eqref{A.16} ensure that the four-divergence (with respect to the Levi--Civita connection associated with the conformal metric $\bar{g}_{ij}$) of the effective energy-momentum tensor
\begin{equation}\label{B.2}
\begin{split}
T_{ij}=\frac{1}{\varphi}\Sigma_{(ij)} - \frac{1}{\varphi^2}V\/(\varphi)g_{ij} + \frac{1}{2\varphi}\left(\hat{T}_hS^h_{\;\;j}U_i + \hat{T}_hS^h_{\;\;i}U_j\right) + \\
+ \frac{1}{2\varphi}\tilde{\nabla}_h\/\left( - S^h_{\;\;i}U_j - S^h_{\;\;j}U_i\right) - \frac{1}{4\varphi^2}S^{pq}S_{pq}U_iU_j + \frac{1}{8\varphi^2}S^{pq}S_{pq}g_{ij}
\end{split} 
\end{equation}
vanishes. To see this point, indicating by $\bar\nabla_i$ the Levi--Civita covariant derivative associated with the conformal metric $\bar{g}_{ij}$, we have 
\begin{equation}\label{B.3}
\begin{split}
\bar{\nabla}^j\/T_{ij}= \frac{1}{\varphi}g^{sj}\bar{\nabla}_s\/T_{ij}= \frac{1}{\varphi}g^{sj}\left[\tilde{\nabla}_s\/T_{ij} - \frac{1}{2\varphi}\left(\de{\varphi}/de{x^i}\delta^q_s + \de{\varphi}/de{x^s}\delta^q_i - \de{\varphi}/de{x^u}g^{uq}g_{si}\right)T_{qj} +\right. \\
\left. - \frac{1}{2\varphi}\left(\de{\varphi}/de{x^j}\delta^q_s + \de{\varphi}/de{x^s}\delta^q_j - \de{\varphi}/de{x^u}g^{uq}g_{sj}\right)T_{iq}\right]
\end{split}
\end{equation}
We have separately
\begin{subequations}\label{B.4}
\begin{equation}\label{B.4a}
\begin{split}
\frac{1}{\varphi}g^{sj}\tilde{\nabla}_s\/T_{ij}=\frac{1}{\varphi}\tilde{\nabla}^j\/\left[\frac{1}{\varphi}\Sigma_{(ij)} - \frac{1}{\varphi^2}V\/(\varphi)g_{ij} + \frac{1}{2\varphi}\left(\hat{T}_hS^h_{\;\;j}U_i + \hat{T}_hS^h_{\;\;i}U_j\right) +\right.\\
\left. + \frac{1}{2\varphi}\tilde{\nabla}_h\/\left( - S^h_{\;\;i}U_j - S^h_{\;\;j}U_i\right) - \frac{1}{4\varphi^2}S^{pq}S_{pq}U_iU_j + \frac{1}{8\varphi^2}S^{pq}S_{pq}g_{ij}\right]=\\
\frac{1}{\varphi^2}\tilde{\nabla}^j\/\left[\Sigma_{(ij)} - \frac{1}{\varphi}V\/(\varphi)g_{ij} + \frac{1}{2}\left(\hat{T}_hS^h_{\;\;j}U_i + \hat{T}_hS^h_{\;\;i}U_j\right) +\right.\\
\left. + \frac{1}{2}\tilde{\nabla}_h\/\left( - S^h_{\;\;i}U_j - S^h_{\;\;j}U_i\right) - \frac{1}{4\varphi}S^{pq}S_{pq}U_iU_j + \frac{1}{8\varphi}S^{pq}S_{pq}g_{ij}\right]+\\
- \frac{\varphi^j}{\varphi^3}\Sigma_{(ij)} + \frac{\varphi^j}{\varphi^4}V(\varphi)g_{ij} - \frac{\varphi^j}{2\varphi^3}\left(\hat{T}_hS^h_{\;\;j}U_i + \hat{T}_hS^h_{\;\;i}U_j\right) +\\
- \frac{\varphi^j}{2\varphi^3}\tilde{\nabla}_h\/\left( - S^h_{\;\;i}U_j - S^h_{\;\;j}U_i\right) + \frac{\varphi^j}{4\varphi^4}S^{pq}S_{pq}U_iU_j - \frac{\varphi^j}{8\varphi^4}S^{pq}S_{pq}g_{ij}
\end{split}
\end{equation}
\begin{equation}\label{B.4b}
\begin{split}
- \frac{1}{2\varphi^2}g^{sj}\left(\varphi_i\/\delta^q_s + \varphi_s\/\delta^q_i - \varphi_u\/g^{uq}g_{si}\right)T_{qj}=\\
- \frac{1}{2\varphi^2}g^{sj}\left(\varphi_i\/\delta^q_s + \varphi_s\/\delta^q_i - \varphi_u\/g^{uq}g_{si}\right)\left[\frac{1}{\varphi}\Sigma_{(qj)} - \frac{1}{\varphi^2}V\/(\varphi)g_{qj} + \frac{1}{2\varphi}\left(\hat{T}_hS^h_{\;\;j}U_q + \right.\right.\\
\left.\left.\hat{T}_hS^h_{\;\;q}U_j\right) + \frac{1}{2\varphi}\tilde{\nabla}_h\/\left( - S^h_{\;\;q}U_j - S^h_{\;\;j}U_q\right) - \frac{1}{4\varphi^2}S^{hk}S_{hk}U_qU_j + \frac{1}{8\varphi^2}S^{hk}S_{hk}g_{qj}\right]=\\
= - \frac{\varphi_i}{2\varphi^3}\Sigma + \frac{4\varphi_i}{\varphi^4}V(\varphi) - \frac{\varphi_i}{8\varphi^4}S_{hk}S^{hk}
\end{split}
\end{equation}
\begin{equation}\label{B.4c}
\begin{split}
- \frac{1}{2\varphi^2}g^{sj}\left(\varphi_j\delta^q_s + \varphi_s\delta^q_j - \varphi_u\/g^{uq}g_{sj}\right)T_{iq}=\\
- \frac{1}{2\varphi^2}g^{sj}\left(\varphi_j\delta^q_s + \varphi_s\delta^q_j - \varphi_u\/g^{uq}g_{sj}\right)\left[\frac{1}{\varphi}\Sigma_{(iq)} - \frac{1}{\varphi^2}V\/(\varphi)g_{iq} + \frac{1}{2\varphi}\left(\hat{T}_hS^h_{\;\;q}U_i + \right.\right.\\
\left.\left. +\hat{T}_hS^h_{\;\;i}U_q\right) + \frac{1}{2\varphi}\tilde{\nabla}_h\/\left( - S^h_{\;\;i}U_q - S^h_{\;\;q}U_i\right) - \frac{1}{4\varphi^2}S^{hk}S_{hk}U_iU_q + \frac{1}{8\varphi^2}S^{hk}S_{hk}g_{iq}\right]=\\
+ \frac{\varphi^j}{\varphi^3}\Sigma_{(ij)} - \frac{\varphi^j}{\varphi^4}V(\varphi)g_{ij} + \frac{\varphi^j}{2\varphi^3}\left(\hat{T}_hS^h_{\;\;j}U_i + \hat{T}_hS^h_{\;\;i}U_j\right) +\\
+ \frac{\varphi^j}{2\varphi^3}\tilde{\nabla}_h\/\left( - S^h_{\;\;i}U_j - S^h_{\;\;j}U_i\right) - \frac{\varphi^j}{4\varphi^4}S^{pq}S_{pq}U_iU_j + \frac{\varphi^j}{8\varphi^4}S^{pq}S_{pq}g_{ij}
\end{split}
\end{equation}
\end{subequations}
Collecting equations \eqref{B.4a}, \eqref{B.4b} and \eqref{B.4c}, we end up with
\begin{equation}\label{B.5}
\begin{split}
\bar{\nabla}^j\/T_{ij}= \frac{\varphi_i}{\varphi^3}\left[-\frac{1}{2}\Sigma + \frac{3}{\varphi}V(\varphi) - V'(\varphi)\right] -\frac{1}{\varphi^2}\left(\frac{\varphi_i}{8\varphi^2}S_{hk}S^{hk}\right) + \\
\frac{1}{\varphi^2}\tilde{\nabla}^j\/\left[\Sigma_{(ij)} - \frac{1}{\varphi}V\/(\varphi)g_{ij} + \frac{1}{2}\left(\hat{T}_hS^h_{\;\;j}U_i + \hat{T}_hS^h_{\;\;i}U_j\right) +\right.\\
\left. + \frac{1}{2}\tilde{\nabla}_h\/\left( - S^h_{\;\;i}U_j - S^h_{\;\;j}U_i\right) - \frac{1}{4\varphi}S^{hk}S_{hk}U_iU_j + \frac{1}{8\varphi}S^{hk}S_{hk}g_{ij}\right]
\end{split}
\end{equation}
Inserting the content of equations \eqref{2.5.13} and \eqref{A.9} into \eqref{B.5}, we conclude that (holding the conservation laws \eqref{2.2}) the requirement $\bar{\nabla}^j\/T_{ij}=0$ is mathematically equivalent to the conditions
\begin{equation}\label{B.6}
- S^{hi}\tilde{\nabla}_h\tilde{\nabla}_iU_j + \frac{1}{2}\tilde{\nabla}_i\tilde{\nabla}_h\left(S^{i}_{\;\;j}U^h\right) - \frac{1}{2}\tilde{\nabla}^i\tilde{\nabla}_h\left(S^h_{\;\;i}U_j + S^h_{\;\;j}U_i\right) =0
\end{equation}
identical to the already obtained equations \eqref{A.16}. The last step (left to the reader) is to verify that, for the conformal metric \eqref{5.1} and the given spin tensor, equations \eqref{A.16} are again trivial identities.  
\\
Repeating the arguments developed in section 4 for the Jordan frame, we can apply again the Bruhat's result \cite{yvonne4} and thus verify that the Hamiltonian constraint \eqref{C.3a} is actually preserved in time.
\section{Examples}
\paragraph{Example 1: $f(R)=R^{\frac{2}{3}}$ model.} The first example we consider is given by the model $f(R)=R^{\frac{2}{3}}$, discussed in the Einstein frame. In this case, the associated potential \eqref{2.5.8} is expressed as
\begin{equation}\label{6.1}
V(\varphi)=-\frac{2}{27}\varphi^{-1}
\end{equation}
while the scalar field \eqref{2.5.7} is given by
\begin{equation}\label{6.2}
\varphi = \frac{4}{3\sqrt{3}}\rho^{-\frac{1}{2}}
\end{equation}
We assume an equation of state for the fluid of the kind $p=\frac{2}{3}\rho$. From equation \eqref{C.6} we have then
\begin{equation}\label{6.3}
\rho = \rho_0\/\left(\bar\tau\Theta\right)^{-\frac{5}{3}}
\end{equation}
Equations \eqref{6.2} and \eqref{6.3} together with the identity $\Theta=\varphi^{-\frac{3}{2}}$ yield the relation
\begin{equation}\label{6.4}
\bar\tau\Theta = \left(\frac{4}{3\sqrt{3}}\right)^{-\frac{2}{3}}{\rho_0}^{\frac{1}{3}}\bar{\tau}^{\frac{4}{9}}
\end{equation}
from which we get the final expressions for the scalar field
\begin{equation}\label{6.5}
\varphi = \left(\frac{4}{3\sqrt{3}}\right)^{\frac{4}{9}}{\rho_0}^{-\frac{2}{9}}\bar{\tau}^{\frac{10}{27}}
\end{equation}
and the energy density
\begin{equation}\label{6.6}
\rho = \left(\frac{4}{3\sqrt{3}}\right)^{\frac{10}{9}}{\rho_0}^{\frac{4}{9}}\bar{\tau}^{-\frac{20}{27}}
\end{equation}
Inserting the content of equations \eqref{6.1}, \eqref{6.5} and \eqref{6.6} into equation \eqref{C.8}, we end up with the final dynamical equation
\begin{equation}\label{6.7}
\ddot{\bar{\tau}} = B\/\bar{\tau}^{-\frac{13}{27}}
\end{equation}
where $B$ is a positive constant (depending on $\rho_0$.) Integrating a first time, we get the relation
\begin{equation}\label{6.8}
\frac{1}{2}\dot{\bar{\tau}}^2= \frac{27}{14}B\bar{\tau}^{\frac{14}{27}} + E
\end{equation}
$E$ being an integration constant. For values $E<0$, equations \eqref{6.7} and \eqref{6.8} imply the existence of a minimum for the volume-scale $\bar\tau$, so giving rise to a universe without singularity. Moreover, by plotting the functions $\bar\tau$ and $\frac{\bar\tau}{t}$, it is seen that the volume-scale $\bar\tau$ is infinite of order $1<n<2$; due to equation \eqref{C.7} this fact implies that the universe isotropizes at large $t$. Also, at large $t$, we have $\frac{{\bar a}^2}{\varphi}\approx\frac{{\bar c}^2}{\varphi}\approx {\bar{\tau}}^{\frac{8}{27}}$ and then the universe expands also with respect to the Jordan frame $g_{ij}=\frac{1}{\varphi}\bar{g}_{ij}$. In connection with this, we notice that the integral $\int_{t_0}^t{\frac{1}{\sqrt{\varphi}}\,dt}$ diverges for $t\rightarrow +\infty$, thus the time reparametrization $d\bar{t}=\frac{1}{\sqrt{\varphi}}dt$ (making the Jordan frame a Bianchi-I metric) is well defined.

\paragraph{Example 2: $f(R)=R^{4}$ model.} We discuss a second example in order to show how solving the field equations in the Jordan or in the Einstein frame are in general two non-equivalent procedures yielding different results. To this end, we consider the model $f(R)=R^4$ coupled with a dust fluid ($\lambda=0$). In such a case, the potential \eqref{2.5.8} is expressed as
\begin{equation}\label{6.13}
V\/(\varphi)= \left(\frac{1}{4}\right)^{\frac{1}{3}}\frac{3}{8}\varphi^{\frac{7}{3}}
\end{equation}
In the Jordan frame, using the first equation of \eqref{4.5}, it is easily seen that the scalar field \eqref{2.5.7} is given by
\begin{equation}\label{6.14}
\varphi = 2^{\frac{5}{4}}\rho_0^{\frac{3}{4}}\tau^{-\frac{3}{4}}
\end{equation}
Inserting equations \eqref{6.13} and \eqref{6.14} into equation \eqref{4.8}, we end up with the final equation 
\begin{equation}\label{6.15}
\frac{1}{4}\ddot\tau = +\frac{3}{16}\frac{{\dot\tau}^2}{\tau} + C\rho_0^{\frac{1}{4}}\tau^{\frac{3}{4}}
\end{equation}
where $C=7(2)^{-\frac{9}{4}}$ is a positive constant. Equation \eqref{6.15} admits solutions of the form 
\begin{equation}\label{6.16}
\tau\/(t) = \left(At+B\right)^8
\end{equation}
with $A$ and $B$ constants. Indeed, substituting equation \eqref{6.16} into equation \eqref{6.15}, we have
\begin{equation}\label{6.17}
\left(2A^2 - C\rho_0^{\frac{1}{4}}\right)\/\left(At + B\right)^6 =0
\end{equation} 
which is always satisfied by suitably choosing the constant $\rho_0$ as function of the constant $A$ or vice versa. From equations \eqref{6.14} and \eqref{6.16} we have $\tau\varphi \approx \left(At+B\right)^2$; due to equation \eqref{4.7} and \eqref{4.7bis}, at large $t$ solutions \eqref{6.16} are seen to isotropize as well as they undergo an accelerated expansion, being $a(t)\approx c(t)\approx \left(At+B\right)^{\frac{8}{3}}$. 
\\
It has not escaped our attention that a model of dust may not necessarily be the most adequate to describe the universe in its early stage; for such a situation, where quantum effects might arise, the coupling to the Dirac field seems to be more appropriate. In this respect, it is worth noticing that in the case of the Dirac field $\psi$ it is possible to find a solution analogous to \eqref{6.16}; indeed in such a circumstance the Dirac field density $\bar{\psi}\psi$ is seen to play the same role of the energy density $\rho$, and the eventual dynamical equation has the same form of \eqref{6.15}. We also stress that in the presence of the Dirac field, the existence of the spin density makes the choice of the initial anisotropy unavoidable; inasmuch as the universe has to evolve according to an initial inflationary expansion together with a phase of isotropization, the solution \eqref{6.16} appears to give a plausible model. For a better understanding of this point, we refer the reader to reference \cite{VFC}.
\\
Now we can study the same model in the Einstein frame. This time we have 
\begin{equation}\label{6.18}
\varphi = 2^{\frac{5}{4}}\rho_0^{\frac{3}{4}}\left(\bar\tau\Theta\right)^{-\frac{3}{4}}
\end{equation}
Making use of the definition $\Theta =\varphi^{-\frac{3}{2}}\/$, from equation \eqref{6.18} we derive the relation
\begin{equation}\label{6.19}
\Theta = 2^{15}\left(\frac{\rho_0}{\bar\tau}\right)^9
\end{equation}
from which we get
\begin{subequations}\label{6.20}
\begin{equation}\label{6.20a}
3\frac{\rho}{\varphi^2} = 3(2)^{12}\rho_0^4{\bar\tau}^{-4}
\end{equation}
\begin{equation}\label{6.20b}
-\frac{6}{\varphi^3}V\/(\varphi) = -9(2)^{4}\rho_0^6{\bar\tau}^{-4}
\end{equation}
\end{subequations}
For values of $\rho_0$ sufficiently small, inserting the content of equations \eqref{6.20} into equation \eqref{C.8} we obtain a final equation of the form
\begin{equation}\label{6.21}
\ddot{\bar\tau}= C\bar{\tau}^{-3} 
\end{equation}
$C$ being a positive constant (dependent on $\rho_0$). Equation \eqref{6.21} can be easily integrated as
\begin{equation}\label{6.22}
\bar\tau = \sqrt{A\/(t+B)^2 + \frac{C}{A}}
\end{equation}
where the constants $A>0$ and $E$ have to be determined consistently with equation \eqref{C.3a}. From equation \eqref{6.22} we have $\bar{\tau}^2 \geq \frac{C}{A}$: we see that in the Einstein frame the model $f(R)=R^4$ coupled to dust can give rise to a universe with no initial singularity but undergoing a slow expansion without isotropization.

\paragraph{Example 3: $f(R)=R^{\frac{3}{2}}$ model.} As third and last example we discuss the model $f(R)=R^{\frac{3}{2}}$ coupled with a fluid having equation of state of the kind $p=\frac{\rho}{2}$. To start with, it is easily seen that the potential \eqref{2.5.8} assumes here the form
\begin{equation}\label{6.9}
V(\varphi)=\frac{2}{27}\varphi^4
\end{equation}
Moreover, from the trace \eqref{2.1bis} of the Einstein-like equations we deduce the relation $R=\rho^{\frac{2}{3}}$ and then from equation \eqref{2.5.7} we have
\begin{equation}\label{6.10}
\varphi = \frac{3}{2}\rho^{\frac{1}{3}}
\end{equation}
Inserting the content of equations \eqref{6.9} and \eqref{6.10} into the dynamical equations \eqref{4.8} and \eqref{C.8}, a direct calculation shows that in both cases the contribution due to density and pressure elides that due to the potential $V(\varphi)$. The conclusion follows that in the Jordan frame (taking the relation $\varphi=\frac{3}{2}\rho_0^{\frac{1}{3}}\tau^{-\frac{1}{2}}$ into account) we end up with the dynamical equation
\begin{equation}\label{6.11}
\ddot\tau = \frac{1}{2}\frac{{\dot\tau}^2}{\tau}
\end{equation}
while in the Einstein frame we have
\begin{equation}\label{6.12}
\ddot{\bar\tau} =0
\end{equation} 
The general (positive) solution of \eqref{6.11} is given by $\tau(t)=\left(At+B\right)^2$, while the solution of \eqref{6.12} is of course $\bar\tau\/(t)=At+B$ ($A$ and $B$ being integration constants). We then see that both solutions in the Jordan and in the Einstein frame have singularity and do not isotropize.
\section{Conclusion}
In the present paper we have studied anisotropic cosmological models of the kind Bianchi type-I arising within the framework of $f(R)$-gravity with torsion coupled with Weyssenhoff spin fluids. Following the lines traced in \cite{VFC} for the coupling with Dirac spinors, we have derived and discussed the field equations in both the Jordan and the Einstein frame, showing that in the latter the Einstein-like equations can be put in a quadratically integrable form. This fact can turn out to be useful for a qualitative analysis of the solutions or numerical integration even in the case analytical integration results hard.
\\
As it has been noticed in previous works \cite{CCSV1,CCSV2,CCSV3,CV4}, in $f(R)$-gravity with torsion there are two possible sources for torsion: the spin density and the non-linearity of the gravitational Lagrangian function $f(R)$. In general, the presence of non-zero spin forces us to take into account anisotropic models, since in such a case torsion does not satisfy the cosmological principle \cite{t}; on the contrary, when spin is zero, the condition of homogeneity gives rise to a torsion consistent with the hypothesis of spatial isotropy, in such a circumstance the torsion being generated by a time-vector with zero spatial components.
\\
Anyway, dealing with Weyssenhoff spin fluids and Bianchi type-I space-times, we have shown that the only significant model is that characterized by spin aligned along a spatial axis and partial isotropy (as it happens in the already known case $f(R)=R$). Moreover, we have proved that for such models the role played by the spin is circumscribed to the assignment of the initial data; indeed, in both the Jordan and Einstein frames, spin contributions are seen to disappear from the dynamical equations \eqref{4.6} and \eqref{4.8} or \eqref{C.8}, while spin is present in the Hamiltonian constraints \eqref{4.3a} and \eqref{C.3a}; the only torsional contribution to the dynamics seems then given by the terms arising from the non-linearity of the function $f(R)$; such terms are present in an identical form also in the case of zero spin. In addition to this, we have studied a few examples, by picking functions of the type $f(R)=R^{n}$ in different cases, and we have seen that the resulting theory seems to be quite sensitive to the specific choice of the model.
\\
Finally, we have derived general conditions assuring the vanishing of the quadridivergence of the Einstein-like equations in both the frames; such conditions turn out to be identically satisfied by the models here considered and allow the preservation in time of the Hamiltonian constraint (momentum constraints are trivial identities); therefore we have proved the correct formulation of the associated Cauchy problem in both the Jordan and Einstein frames.

\end{document}